\documentclass[prl,twocolumn,superscriptaddress]{revtex4-1}
\usepackage{amsmath, amssymb,graphicx,color,bm,epstopdf}
\usepackage[dvipsnames]{xcolor}
\usepackage{bbold}
\usepackage{braket}
\usepackage{comment}

\begin{document}

\title{Reply to “Comment on ``Quantum Time Crystals from Hamiltonians with Long-Range Interactions''"}

\author{Valerii K. Kozin}
\affiliation{Science Institute, University of Iceland, Dunhagi 3, IS-107, Reykjavik, Iceland}
\affiliation{ITMO University, Kronverkskiy prospekt 49, Saint Petersburg 197101, Russia}

\author{Oleksandr Kyriienko}
\affiliation{Department of Physics and Astronomy, University of Exeter, Stocker Road, Exeter EX4 4QL, UK}
\affiliation{Department of Nanophotonics and Metamaterials, ITMO University, St. Petersburg, 197101, Russia}

\begin{abstract}
In the note by Khemani et al. [arXiv:2001.11037] the authors express conceptual disagreement with our recent paper on quantum time crystals [Phys. Rev. Lett. 123, 210602]. They criticise the idealized nature of the considered quantum time crystal, and make several points about properties of Hamiltonians presented in our work. In this reply we answer one-by-one all questions raised in the discussion. As for the ideological dispute, it brightly highlights a bizarre nature of time crystalline order in closed quantum systems, and we offer a different vision for the development of the field.
\end{abstract}

\maketitle

In the paper~\cite{KozinKyriienko2019} we discuss the time crystalline behavior following a well-accepted definition by Watanabe and Oshikawa~\cite{WatanabeOshikawa2015}. It concerns a \emph{closed} quantum system of coupled spins where drive and decay processes are absent (as envisioned by Frank Wilczek in the outlook of Ref.~\cite{Wilczek2012}). Searching for a suitable Hamiltonian that allows for time-periodic spin correlations to exist in the thermodynamic limit, we discover a family of time crystal (TC) Hamiltonians satisfying this condition, yet having diverse properties. The characteristic trait of TC Hamiltonians is the presence of multiparticle interaction which involves $O(N)$ spins, being the consequence of Watanabe-Oshikawa definition, that can be traced to a demand that oscillations of magnetization correlation function remain non-zero in the limit of infinite number of particles $N$.

First, in the note Khemani et al.~\cite{khemani2020comment} criticise the considered system for being ideal and compare it to the mathematical pendulum concept. We show that this is not the case. If as suggested we assume that the center of mass coordinate $\hat{x}_{\text{CM}}=N^{-1}\sum_i^N \hat{x}_i$ and momentum $\hat{p}_{\text{CM}}=\sum_i^N \hat{p}_i$ of a pendulum are isolated from other degrees of freedom, the resulting Hamiltonian has the form\linebreak $\hat{H}_{\text{CM}}=\hat{p}_{\text{CM}}^2/(2N m_0)+N m_0\omega^2\hat{x}_{\text{CM}}^2/2$,
where $N m_0$ is a total mass of the system, and $\omega$ is its frequency. By direct calculation one can check that the ground state correlation function \linebreak $\langle0\lvert\hat{x}_{\text{CM}}(t)\hat{x}_{\text{CM}}(0)\rvert0\rangle=\hbar/(2N m_0\omega)e^{-i\omega t}$ vanishes in the thermodynamic limit $N\rightarrow\infty$, making it unsuitable for the role of a time crystal.

Second, Khemani et al. suggest introducing $N$-body observables, stating that in this case every system is a time crystal. We note that the claim does not apply to our system as we strictly follow the conventional definition by Watanabe and Oshikawa~\cite{WatanabeOshikawa2015}. If we are allowed to change definitions, time crystalline behavior becomes possible with finite-range interactions~\cite{IsolatedHeisenbergTC}. 

Next, to start the analysis of possible TC Hamiltonians in~\cite{KozinKyriienko2019} we present several toy models described by Eqs.~(5) and (9). They serve a solely \emph{pedagogical} purpose. The concern about stability and simplicity of these Hamiltonians is not crucial to understanding the concept. They are introduced as intermediate steps on the way towards one of the main results of the paper -- the Hamiltonian in Eqs. (10)--(11). It reads $\hat{H}(J) = -\sum\limits_{j=1}^N\sigma_z^{(j)}\sigma_z^{(j+1)}+
    J\big(\sigma_x^{(1)}\sigma_x^{(2)}...\sigma_x^{([N/2])}-\sigma_x^{([N/2]+1)}...\sigma_x^{(N)} \big)$,
where $\sigma_z^{(j)}$ and $\sigma_x^{(j)}$ are standard Pauli operators acting at site $j$. This Hamiltonian, first introduced in Ref.~\cite{FlorioPRL2011} outside of the context of time crystals, is stable at zero temperature against perturbations such as Heisenberg exchange or a random onsite field and has non-trivial many-body Hilbert space, thus removing the claims in Ref.~\cite{khemani2020comment}. We further note that Hamiltonian $\hat{H}(J)$ is one of many possible TC Hamiltonians defined by conditions (a) and (b) (p. 2 in Ref.~\cite{KozinKyriienko2019}), and is chosen as the one hosting nondegenerate GHZ states in the spectrum.

The finishing part of \cite{khemani2020comment} notes that TC Hamiltonians discovered in the closed quantum system are reminiscent to Floquet time crystals governed by a simple drive sequence, yet require long-range multiparticle interaction. This is exactly the point we consider in the end of our paper~\cite{KozinKyriienko2019}, and no new information is presented.

Next, Ref.~\cite{khemani2020comment} highlights an infinite range of proposed TC Hamiltonians and questions the influence of retardation. While quantum TC Hamiltonian $\hat{H}(J)$ is physically reasonable, as explained above, we agree that it is an extremely complicated Hamiltonian to simulate for increasing number of sites. In fact, the original no-go theorem from Ref.~\cite{WatanabeOshikawa2015} was recently strengthened~\cite{huang2019absence, Watanabe2020} to show that any Hamiltonian, represented by a sum of finite range terms with a constant-bounded operator norm, can not represent a quantum time crystal. 
Thus, if one manages to improve the locality of TC Hamiltonian as compared to the Hamiltonian $\hat{H}(J)$ with $N/2$-body couplings---an interesting quest by itself---then in according to aforementioned theorems it would inevitably contain infinite-range terms in the thermodynamic limit. Recently we became aware of the study~\cite{EfetovTC2019}, where it is claimed that one may realize a time crystal using infinite-range, but finite-body interactions. 

We note that infinite range $m$-body interactions in $m\le N$-body systems are actively studied in condensed matter physics, quantum field theory~\cite{SachdevYe,Kitaev,RosenhausSYK_rev,genSYK} and quantum information~\cite{graphStates2008,Babbush2014}. They are related to Hamiltonians usually simulated in quantum chemistry~\cite{Cao2019,Cao2019,Babbush2014}, as Jordan-Wigner transformation of fermionic Hamiltonians generally leads to qubit (spin-1/2's) Hamiltonians in the form of spin string sums. Such Hamiltonians with multi-particle interactions appear in systems where some of degrees of freedom are integrated out~\cite{Babbush2014}. The requirement for an ideal noiseless operation thus poses the question: do we need fault-tolerant quantum devices to realize quantum time crystals?


\end{document}